\documentclass[twoside,fleqn]{article}
\usepackage{espcrc2,graphicx,here}
\newcommand{\beq}{\begin{equation}}
\newcommand{\eeq}{\end{equation}}
\newcommand{\beqa}{\begin{eqnarray}}
\newcommand{\eeqa}{\end{eqnarray}}
\newcommand{\beqan}{\begin{eqnarray*}}
\newcommand{\eeqan}{\end{eqnarray*}}
\newcommand{\ba}{\begin{array}}
\newcommand{\ea}{\end{array}}

\newcommand{\nn}{\nonumber \\}
\newcommand{\bea}{\begin{eqnarray}}
\newcommand{\eea}{\end{eqnarray}}

\newcommand{\hepph}[1]{{\tt hep-ph/#1}}

\newcommand{\AmS}{{\protect\the\textfont2
   A\kern-.1667em\lower.5ex\hbox{M}\kern-.125emS}}

\title{
\vspace{-1.0cm}
{\sf \small \rightline{UWThPh-2002-26}}
\bigskip
Four-pion production\thanks{Work supported in part by TMR, 
EC-Contract  No. ERBFMRX-CT980169 (EURODA$\Phi$NE).}}

\author{G. Ecker and R. Unterdorfer\address{Institut f\"ur 
Theoretische Physik, Universit\"at Wien\\ Boltzmanngasse 5, 
A-1090 Vienna, Austria} }

\begin{document}

\begin{abstract}
Starting from the low-energy structure derived from
QCD, we extend the amplitudes for four-pion production in 
$e^+ e^-$ annihilation and $\tau$ decays up to invariant 
four-pion masses of
1 GeV. Cross sections and branching ratios 
$BR(\rho^0 \to 4 \pi)$ are compared with available data.
\end{abstract}

\maketitle

\section{INTRODUCTION}

The production of four pions in $e^+ e^-$ annihilation and in $\tau$ 
decays is interesting in its own right but it also represents a
non-negligible component of hadronic vacuum polarization. In fact, 
almost 5 $\%$ of the lowest-order hadronic contribution to the 
anomalous magnetic moment of the muon $a_\mu$ is due to four pions
\cite{dehz02}. At the level of accuracy required for a comparison of 
the standard model prediction for $a_\mu$ with the recent BNL
measurement \cite{bnl02}, better knowledge of this contribution would
be welcome (see the discussion in Ref.~\cite{dehz02}). For the
determination of the fine structure constant at $s=M_Z^2$, the
relative importance of the four-pion contribution is even bigger but
in this case the available precision is sufficient for the time being.

In this talk, I report on the work of Ref.~\cite{eu02} where we
have constructed the relevant amplitudes up to invariant four-pion
masses of about 1 GeV. This is not enough for the purpose of
calculating $a_\mu$, but it is a first step in this direction. Even if
it is impossible to calculate the amplitudes directly from QCD 
our aim was to construct amplitudes that are at least consistent with 
QCD. This program turned out to have some surprises in store. 

The procedure of Ref.~\cite{eu02} starts from a calculation to 
next-to-leading order in the low-energy expansion of QCD, employing the
methods of chiral perturbation theory (CHPT) \cite{glchpt}. The
corresponding low-energy amplitudes cannot be used directly in the 
physical region because the four-pion threshold of 560 MeV is already
close to the resonance region. But the low-energy amplitudes
contain nontrivial information how to continue to higher energies. At
next-to-leading order, the traces of $\rho$ and scalar meson exchange
appear in the amplitudes. Supplemented by $\omega$, $a_1$ and double
$\rho$ exchange, the resulting $e^+e^-$ cross sections describe
the available experimental data very well up to cms energies of about 
1 GeV.

\section{SYMMETRIES AND LOW-ENERGY LIMIT}

There are altogether four different channels accessible in 
$e^+ e^-$ annihilation and $\tau$ decays into four pions. In the
isospin limit that we assume throughout, the amplitudes of either
$e^+ e^- \to 2 \pi^0 \pi^+ \pi^-$ \cite{kuehnrev} or
$\tau^- \to \nu_\tau ~2 \pi^- \pi^+ \pi^0$ \cite{eu02} are sufficient
to determine all four amplitudes. One important advantage of the
chiral approach
is that not only chiral symmetry but also charge conjugation 
invariance, Bose symmetry and electromagnetic gauge invariance are
manifest at each stage of the calculation.

\begin{figure}[hbt]
\begin{center}
%\hspace*{-0.5cm} 
\includegraphics[width=7cm]{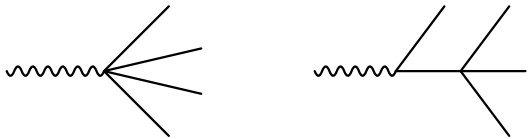}
\caption{Tree diagrams for $\gamma^* \to 4 \pi$. Solid lines denote 
pions and the wavy line stands for the virtual photon.}
\label{fig:tree}
\end{center}
\end{figure}

At leading order in the low-energy expansion, the amplitude is
completely determined by ``virtual'' bremsstrahlung. From the diagrams
in Fig.~\ref{fig:tree}, the matrix element of the electromagnetic 
current governing the amplitude for $e^+ e^- \to 2 \pi^0 \pi^+ \pi^-$
is found to be
\begin{eqnarray}  
\langle \pi^0(p_1) \pi^0(p_2) \pi^+(p_3) \pi^-(p_4)
|J^\mu_{\rm elm}(0)|0\rangle &= & \nn
\displaystyle\frac{s-M_\pi^2}{F_\pi^2}
\left(\displaystyle\frac{2 p_3^\mu}{2 p_3\cdot q - q^2} -
\displaystyle\frac{2 p_4^\mu}{2 p_4\cdot q - q^2}\right)~, &&
\label{eq:tree}
\end{eqnarray} 
where $s=(p_1+p_2)^2$, $q=p_1+p_2+p_3+p_4$ and $F_\pi=$
92.4 MeV is the pion decay constant. This matrix element has a very
suggestive structure: $(s-M_\pi^2)/F_\pi^2$ is the (lowest-order) 
scattering amplitude for 
$\pi^0 \pi^0 \to \pi^+ \pi^-$ and the second term reduces to the
usual bremsstrahlung factor for real photons ($q^2 \to 0$).

Although the amplitude (\ref{eq:tree}) is by itself not of direct
phenomenological relevance we can use the isospin relations
\cite{kuehnrev} and calculate also the leading-order $\tau$ decay
amplitudes. In the chiral limit ($M_\pi=0$), those amplitudes should
coincide with those of Ref.~\cite{fww80}. Unfortunately, there are two
(identical) misprints in Ref.~\cite{fww80} that have moreover
propagated into some of the subsequent literature (\cite{ck01} and
references therein). Although the structure is correct, the
normalization is not: the correct amplitude (\ref{eq:tree}) and the
corresponding $\tau$ decay amplitudes are smaller by a factor
$\sqrt{2}/(3\sqrt{3})$ (or $1/13.5$ in rate). This normalization error
also affects the so-called CLEO current \cite{dfhj96} in the Monte
Carlo package TAUOLA \cite{tauola}. 

The ``current algebra'' amplitude (\ref{eq:tree}) is important for 
checking the low-energy limit of QCD but it is not a realistic
approximation in the physical region. In order to see the traces of
meson resonance exchange, we have to go at least to
next-to-leading order.

\section{RESONANCE EXCHANGE}

At next-to-leading order in the chiral expansion, the amplitude
consists of two parts: a loop amplitude \cite{unter02}
and a tree amplitude containing 
the (renormalized) coupling constants of the chiral Lagrangian of 
$O(p^4)$ \cite{glchpt}. With the standard values of those constants,
one arrives at cross sections that are still unrealistic. As indicated
by the dotted curves in Figs.~2,3, the theoretical cross sections are
significantly smaller than the measured ones.

The seeds of meson resonance exchange appear first in the coupling 
constants of $O(p^4)$. In fact, those constants are known to be 
saturated  by meson resonance exchange to a large extent
\cite{egpr89,eglpr89}. This saturation makes the matching between the 
strictly chiral amplitude to $O(p^4)$ and a more realistic meson
resonance exchange amplitude almost
trivial. Using the standard chiral resonance Lagrangian \cite{egpr89},
the resonance amplitudes are guaranteed to exhibit the correct
low-energy behaviour to $O(p^4)$.

In four-pion production, only the $\rho$ and the (isoscalar) scalar
mesons contribute at $O(p^4)$. As could have been expected, $\rho$ 
exchange dominates by far. The overall contribution from scalar exchange
turns out to be very small so that the controversial structure of the
scalar sector is not relevant in practice.

The modified amplitudes with $\rho$ and scalar exchange are
definitely more realistic than the chiral low-energy amplitudes.
However, except in the vicinity of the $\rho$ pole, the resulting
cross sections are still too small (not shown in Figs.~2,3). The
obvious lesson is that important ingredients of the amplitudes 
are still missing that only show up at orders $p^6$ or higher in 
the chiral expansion.

The most important missing degrees of freedom are easily found: both
quantum number considerations and experimental information
\cite{cmdh,cmdl} indicate that $\omega$ and $a_1$ exchange must be
incorporated. Whereas the lowest-order coupling of the $\omega$ to
pions is unique, there is some ambiguity in the $a_1\rho\pi$ couplings
(to be resolved eventually by studies of three-pion production 
\cite{pp02}).
With a simplifying assumption for those couplings \cite{eu02},
including double $\rho$ exchange that also comes in at $O(p^6)$ and
performing a resummation of some terms making up the $\rho$-dominated pion
form factor, the amplitudes assume their final form. Except for the
ambiguity in the $a_1\rho\pi$ couplings, all resonance couplings can be
determined from the respective decay widths.

\section{COMPARISON WITH DATA}

The two main assets of our amplitudes are:
\begin{itemize} 
\item They contain the relevant degrees of freedom for describing
four-pion production up to energies of about 1 GeV.
\item They exhibit the correct low-energy behaviour to $O(p^4)$ by
construction.
\end{itemize} 

For energies below 1 GeV, annihilation data are available for the 
channel $2 \pi^+ 2 \pi^-$ mainly. In Fig.~2, the theoretical cross 
sections are compared with the most recent (and most precise) data 
from the CMD-2 Collaboration \cite{cmdl} (see Ref.~\cite{eu02} for 
the full data set). The cross section for our model is shown as the 
full curve. The dashed curve corresponds to omitting the loop
amplitude of $O(p^4)$ \cite{unter02}
(except for the contribution to the width of the 
$\rho$ meson). The obvious conclusion is that the amplitude is completely 
dominated by resonance exchange. Although a possible enhancement of
the cross section in the region between 800 and 900 MeV could not be 
explained with our amplitude the gross features 
of the data can be reproduced over a range of two orders of 
magnitude with almost no free parameters. In retrospect, the
simplifying assumption for the $a_1\rho\pi$ couplings (that actually
has a theoretical basis \cite{pp02}) is justified by comparison
with experiment: other
choices for the couplings could not reproduce the data. Note that 
$\omega$ exchange does not contribute in this channel: the cross section
near 1 GeV is completely dominated by $a_1$ exchange.

The experimental situation is less satisfactory for the other
annihilation channel $2 \pi^0 \pi^+  \pi^-$. The most precise
experiment \cite{cmdh} has measured the cross section at only two
energies below 1.05 GeV. The comparison between theory and experiment
is shown in Fig.~3. The loop contribution is even less relevant in
this case. The theoretical cross section increases by three orders of
magnitude from the $\rho$ resonance to match the two data
points. For this channel, the cross section  near 1 GeV is 
dominated by $\omega$ exchange. The theoretical prediction is
therefore less sensitive to assumptions about the $a_1\rho\pi$ 
couplings.

Our amplitudes and the corresponding cross sections cannot be extended
to the phenomenologically most interesting region above 1 GeV without
further input. The reason is that the amplitudes do not satisfy the 
high-energy constraints of QCD. In fact, the theoretical cross
sections exceed the data soon above 1 GeV of cms energy. Additional
higher-mass states must be included to access the region up
to 2 GeV and to ensure a proper high-energy behaviour. Resummations 
similar to the pion form factor may also be necessary.

\begin{figure}[hbt]
\begin{center}
\includegraphics[width=7.5cm]{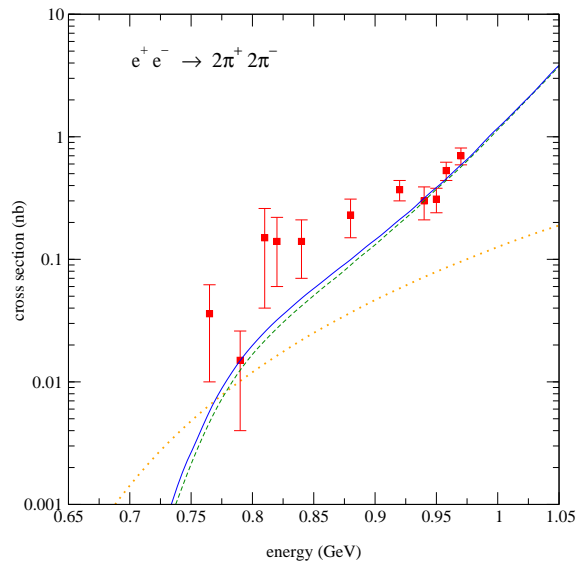}
\caption{Comparison of theoretical predictions for $e^+ e^- \to 
2\pi^+ 2\pi^-$ with data \cite{cmdl}: complete cross section (full
curve), without loop contributions (dashed curve), cross section for
the amplitude of $O(p^4)$ (dotted curve).}
\end{center}
\label{fig:cc}
\end{figure}

\begin{figure}[hbt]
\begin{center}
\includegraphics[width=7.5cm]{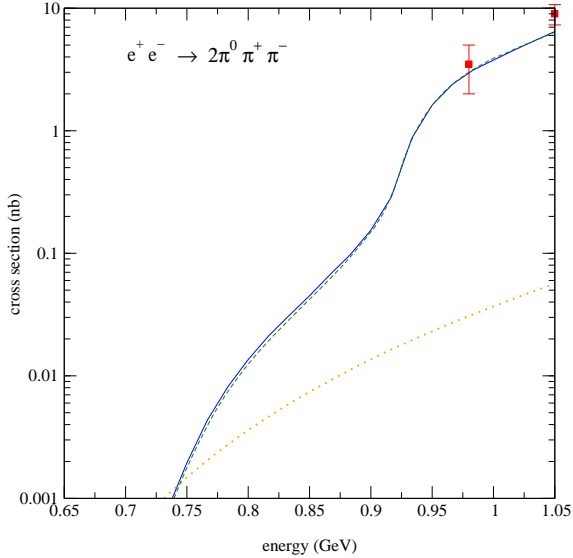}
\caption{Same as in Fig.~2 for the process $e^+ e^- \to 2 \pi^0
\pi^+ \pi^-$ \cite{cmdh}.}
\end{center}
\label{fig:nc}
\end{figure}

From our amplitudes we can also extract the branching ratios for the
four-pion decay modes of the $\rho^0$. For $q^2=M_\rho^2$ several
contributions to the amplitudes are of comparable size, with partly 
destructive interference. In this way, uncertainties in the resonance 
couplings are enhanced. We therefore quote predictions for the 
branching ratios with a 40 $\%$ uncertainty:
\begin{eqnarray} 
BR(\rho^0 \to 2 \pi^+ 2\pi^- ) &=& (6.7 \pm 2.7)\times 10^{-6} \\
BR(\rho^0 \to 2 \pi^0 \pi^+ \pi^- ) &=& (5.0 \pm 2.0)\times 
10^{-6}.
\end{eqnarray} 
For comparison, the Particle Data Group \cite{pdg00} lists 
\begin{eqnarray} 
BR(\rho^0 \to 2 \pi^+ 2\pi^- ) &=& (1.8 \pm 0.9)\times 10^{-5} \\
BR(\rho^0 \to 2 \pi^0 \pi^+ \pi^- ) &<& 4 \times 10^{-5}.
\end{eqnarray}

\section{CONCLUSIONS}

The following features of our model for four-pion
production are worth repeating:
\begin{itemize} 
\item The amplitudes exhibit the correct low-energy structure to $O(p^4)$
in the chiral expansion.
\item All symmetries of the transitions are manifest in the QFT
framework of CHPT: (broken) chiral symmetry, gauge invariance, Bose
symmetry and charge conjugation.
\item In addition to $\rho$ (and the less important scalar) 
exchange, $\omega$ and $a_1$ exchange are crucial for understanding 
the experimental results already at energies below 1 GeV.
\item Good agreement with available data is obtained, covering
several orders of magnitude in cross sections.
\item To extend the amplitudes to energies above 1 GeV,
the correct high-energy behaviour still needs to be implemented.
For the same reason, comparison with $\tau$ decay data is postponed.
\end{itemize}

\end{document}